\documentclass[doublecol]{epl2} 

\usepackage{subfigure}
\usepackage[toc,page]{appendix}

\newcommand*{\GeV}{\ifmmode {\mathrm{\ Ge\kern -0.1em V}}\else
                   \textrm{Ge\kern -0.1em V}\fi}%
\newcommand*{\TeV}{\ifmmode {\mathrm{\ Te\kern -0.1em V}}\else
                   \textrm{Te\kern -0.1em V}\fi}%

\def\pT{\ensuremath{p_{\mathrm{T}}}}

\title{Inclusive production of X$\rightarrow b\bar{b}$ plus a recoil for the LHC Run-II}
\shorttitle{Inclusive production of X$\rightarrow b\bar{b}$ plus a recoil for the LHC Run-II} 

\author{Nicolas Gutierrez\inst{1,2}}
\shortauthor{N. Gutierrez \etal}

\institute{                    
  \inst{1} Department of Physics and Astronomy, University College London - London, WC1E 6BT, United Kingdom\\
  \inst{2} Institute for Particle Physics Phenomenology, Department of Physics, Durham University - DH1 3LE, United Kingdom
}
\pacs{aa/bb/cc}{First pacs description}
\pacs{aa/bb/cc}{Second pacs description}

\abstract{
  This letter presents a study of the inclusive production of X$\rightarrow b\bar{b}$ 
  plus a recoil, using simulated samples of $pp$ collisions at $\sqrt{s}=14$ \TeV\ 
  for an integrated luminosity in the range between 30 fb$^{-1}$ to 3 ab$^{-1}$.
  The case for experiments to include un-prescaled $b$-tag multijet triggers for this topology
  is made and the ideal jet thresholds are discussed.
  The sensitivity to Standard Model Higgs with a transverse
  momentum of at least 200~\GeV\ is evaluated with respect to a continuous background, dominated
  by multijet processes. 
  The mass of $b$-jet-pairs is analysed, quoting sensitivity to cross-sections in the range
  of 1 to 2~pb, for 100 fb$^{-1}$,
  covering the Higgs production cross section of 1.8~pb.
  The trigger strategy presented in this letter is compared to triggers already in use,
  showing an increase on the signal efficiency for masses below 200~\GeV\ and a performance comparable
  to a logic OR of all the currently available akin triggers for higher masses.
  The robustness of the expected sensitivity against systematic uncertainties is estimated by considering 
  various typical sources, such as those on the fitting parameters of the continuous background, 
  shape uncertainties affecting the signal acceptance and the background modelling.
  The accuracy of a Higgs production cross section measurements is also discussed,
  quoting sensitivity to deviations of 50\%\ for 100~fb$^{-1}$ and 10\%\ for 3~ab$^{-1}$.
}
\begin{document}

\maketitle

\section{Introduction}

Although the Higgs boson has been discovered at the LHC~\cite{Aad:2012tfa,Chatrchyan:2012ufa}, 
a $5\sigma$ observation for  its largest decay mode $H \rightarrow b\bar{b}$  has not been achieved so far.
The combination of huge backgrounds with large modelling uncertainties transform the observation of this
hadronic decay mode in a non-trivial task. The usual production channels, Higgs-Strahlung  $(VH)$, associated top-pair Higgs 
$t\bar{t}H$ and Vector Boson Fusion (VBF), provide upper limits on the event rate of respectively $1.4$, $1.5$ and $2.8$ times the
Standard Model (SM) expectation~\cite{Chatrchyan:2013zna,Aad:2015gra,Khachatryan:2015bnx}. 
Measuring the Higgs-bottom coupling is of crucial importance at the LHC. 
Not many channels to measure this coupling exist. So far only $HZ$, 
$ttH$ and VBF have been studied~\cite{Butterworth:2015bya,Soper:2010xk,Englert:2015dlp,Moretti:2015vaa,Plehn:2009rk}. 
In this letter, a new channel in gluon fusion is explored.

The early observation of $H \rightarrow \gamma\gamma$ at the LHC demonstrates that
despite a small signal to background ratio, a data-driven background approach, high statistics and control over
systematic uncertainties can yield a discovery. This letter explores the possibility of pursuing this strategy for the
SM $H \rightarrow b\bar{b}$ channel, as well as for the search for SM $Z$-bosons and Beyond the SM CP-odd (A)
bosons decaying into $b\bar{b}$ predicted by the two-Higgs-doublet model (2HDM).

The study presented here explores the boosted $X\rightarrow b\bar{b}$ kinematic regime such that the combinatorics 
are reduced, yet each heavy resonance $X$ decay can still be resolved via standard jet-clustering algorithms with typical jet-radius.
Requiring high-\pT\ jets is further motivated by the limited capability of trigger systems of detectors to record events with a high rate.
This, however, can be compensated by triggering on events with multiple $b$-jets, which is the strategy explored in the following. The 
dominant multi-jet background comes from the pure QCD $b\bar{b}$+jets production. Additional $b$-jets from $g \rightarrow b\bar{b}$ may be 
present in both signal and background - all such candidates are included in the analysis. 
Interestingly, the  $Z \rightarrow b\bar{b}$ channel has 
already been measured using the 8 \TeV\ data with a similar analysis strategy~\cite{Aad:2014bla}.

\begin{figure*}[!t]
  \subfigure[$R=0.4$]{\includegraphics[width=0.49\textwidth]{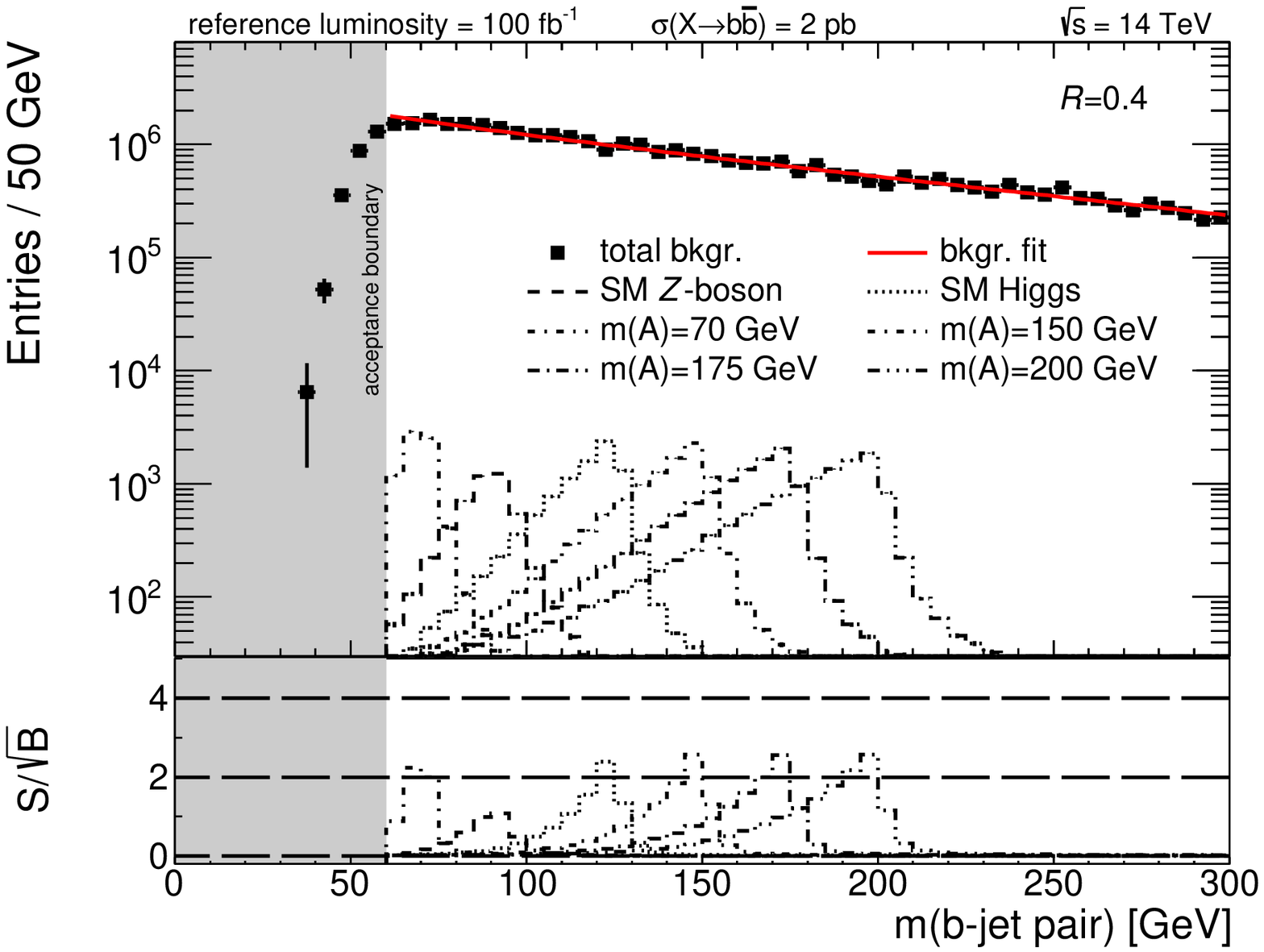}}
  \subfigure[$R=0.2$]{\includegraphics[width=0.49\textwidth]{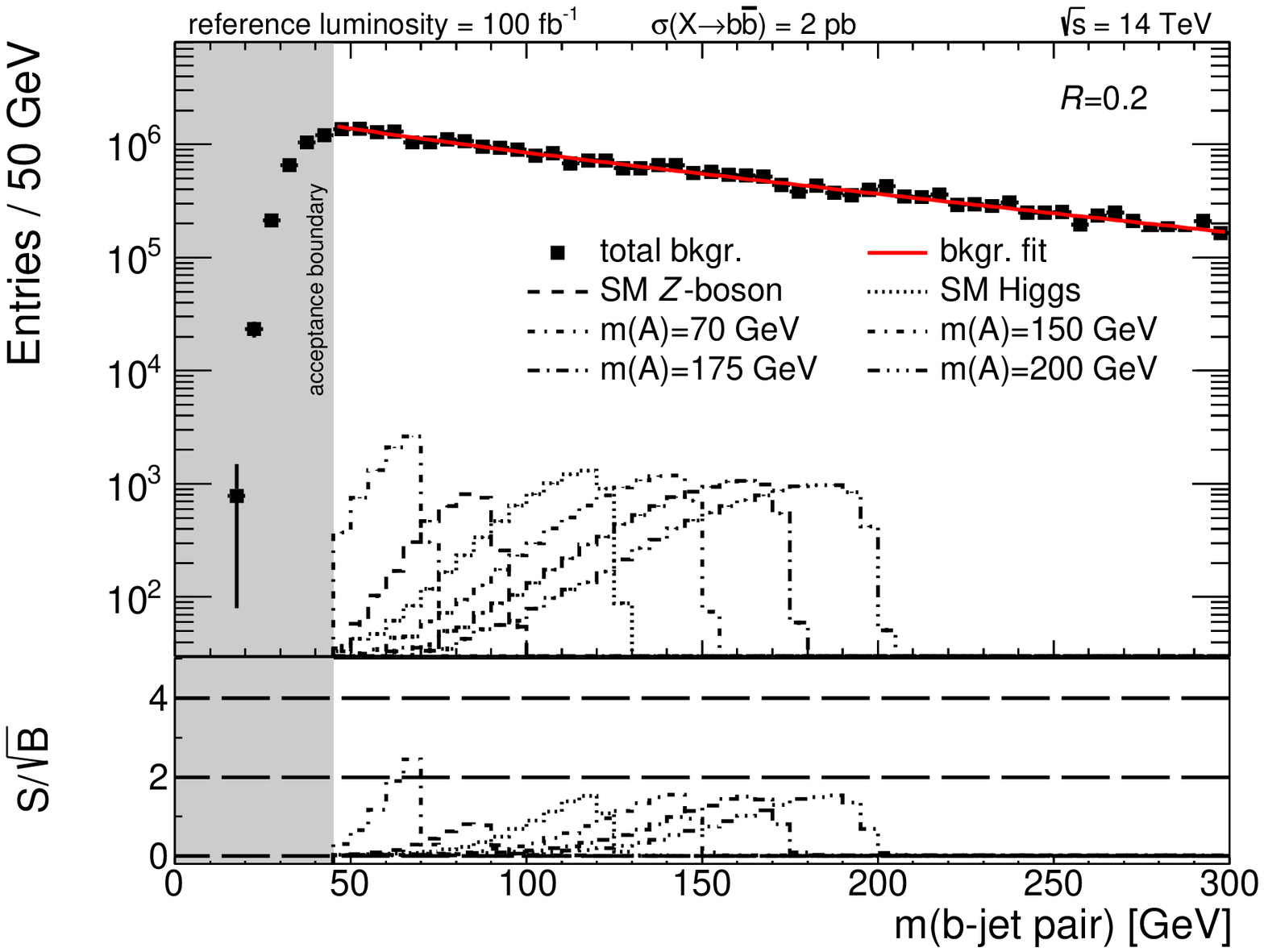}}
  \caption{Invariant $b\bar{b}$ mass distribution for the signal hypothesis and total background contribution.
  Signal samples with masses between 70 and 200~\GeV\ are shown.
  The total background is fitted to the dijet mass shape shown as the continuous line (red).
  The bottom panel displays the ratio between the signal yield and the expected statistical uncertainty
  on the background.  A cross-section of 2~pb for all the signal samples is used.}
  \label{fig:candidatemass} 
\end{figure*}

This letter is organised as follows: First, the analysis strategy is presented, devoting special emphasis to the 
background fitting approach. Second, the LHC Run II bounds to the SM Higgs and Z
bosons via the strategy proposed here are derived. 

\section{Event Generation and Analysis}
\label{sec:analysis}

Signal Monte Carlo (MC) samples are generated for scalar and pseudo-scalar production with
\textsc{MCFM+Pythia8}~\cite{ref:MCFM,Sjostrand:2007gs}  accounting for the full top and bottom mass 
effects in the production vertex. The higher order QCD effects are included with a flat correction factor of 
$K\sim1.6$~\cite{Campbell:2012am,Buschmann:2014twa}. For the $Z$-boson signal and dijets and 
$t\bar{t}$+jets backgrounds, the~\textsc{Sherpa} event generator in the four-flavour scheme~\cite{sherpa} is used.
All spin correlation effects in the $Z$ and top decays are accounted for. 
These samples are merged up to two extra jets via the \textsc{CKKW} 
algorithm~\cite{Catani:2001cc,Hoeche:2009rj}. 

The simulated events are analysed using \textsc{Rivet}~\cite{Buckley:2010ar}, 
which makes extensive use of~\textsc{fastjet}~\cite{Cacciari:2008gp,Cacciari:2011ma} for the jet finding.
Jets are defined with the anti-$k_{t}$  algorithm with a radius $R=0.4$ satisfying $\pT>30$ \GeV\ and $|\eta| < 3.0$.
To be able to reconstruct the resonance $X\rightarrow b\bar{b}$ and suppress the combinatoric background, 
two b-tagged jets are required.
A $b$-tagging efficiency of 70\%\ and 1\% of miss-tagging rate~\cite{Aad:2015ydr} are assumed.

Any event with an isolated electron or muon is rejected. A lepton is isolated if the hadronic transverse  energy within a cone
of $R=0.2$ surrounding the lepton tantamount  to less than  20\% of the lepton transverse energy deposited.
Boson-candidates are reconstructed by adding the four-vectors of the two b-tagged jets. Each candidate is required to have a minimum
transverse momentum of $\pT > 200$~\GeV.  \medskip


\begin{figure}[!b]
  \includegraphics[width=0.49\textwidth]{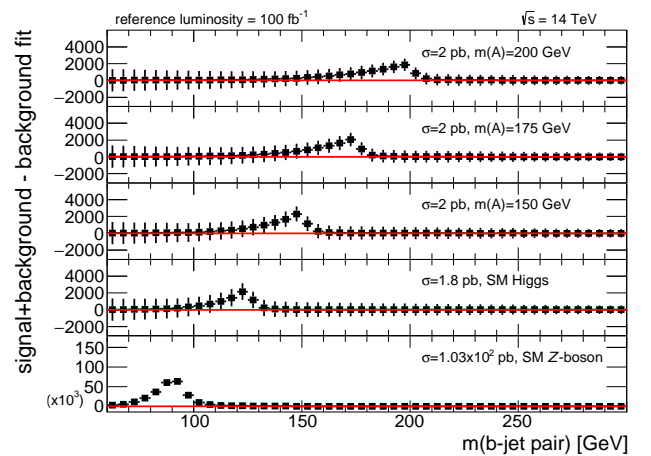}
  \caption{Difference between the signal-injected background and the background fit for the SM Higgs and $Z$ bosons, Higgs 
  and the pseudo scalar  boson $m_A=150,~175,~200$~\GeV~signal samples.
  The signal injection is performed and shown for each signal sample separately.}
  \label{fig:sigOverStat} 
\end{figure}

Fig.~\ref{fig:candidatemass}a shows the invariant mass distribution for the Higgs, $Z$ and pseudo-scalar $A$ signals 
and total background after the event selection. The background is fitted to a functional form
\begin{equation}
  p_{1}(1-x)^{p_{2}}x^{p_{3}}.
\label{eq:FitForm}
\end{equation}
An equivalent function is used in similar searches for resonances against a steeply falling background
in hadronic final states~\cite{Aad:2015owa,Khachatryan:2015bma}.
Also in this Fig., all the signal cross-sections are set to 2~pb. 
A reference luminosity of 100 fb$^{-1}$ is used.
The bottom panel illustrates the expected sensitivity, by computing a bin-by-bin ratio
of the number of expected signal events to the statistical uncertainty on the background prediction. 
The grey shaded area illustrates the lower bound of the fitting range.
This boundary is caused by the $R=0.4$ jets no longer being able to resolve each b-hadron as an individual jet.
This can be illustrated  by running the event selection with a smaller jet-radius 
parameter $R=0.2$, see Fig.~\ref{fig:candidatemass}b.
Here, the reach of the fitted form for low masses increases by 15~\GeV.
The usage of a smaller $R$-parameter, however, also increases the out-of-cone radiation,
therefore degrading the signal mass resolution.
Techniques to better resolve the collimated $b\bar{b}$ system usually rely on jet-substructure 
technics~\cite{Butterworth:2008iy,Krohn:2009zg,Thaler:2015xaa}. However, this is outside of the scope 
of this paper as most of the signal events are in the resolved regime for our considered mass points.\medskip

Fig.~\ref{fig:sigOverStat} shows the difference between signal-injected background 
and the background fit using the generator cross-section predictions for the SM Higgs and Z bosons,
and a fixed 2 pb cross-section for CP-odd bosons for a reference luminosity of 100 fb$^{-1}$.
This illustrates the size of the expected signal bump with respect to the statistical
uncertainty on the background prediction.\medskip

\subsection{Trigger effects}

\begin{figure}[!b]
  \includegraphics[width=0.49\textwidth]{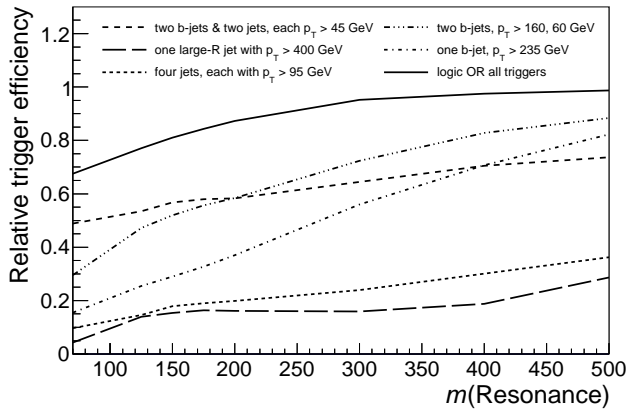}
  \caption{Efficiency of various triggers and for the combination of all these triggers.}
  \label{fig:TrigEff} 
\end{figure}

\begin{figure}[!b]
  \includegraphics[width=0.49\textwidth]{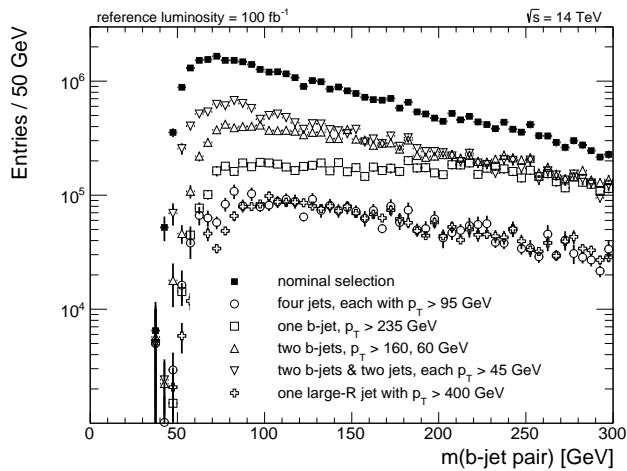}
  \caption{Total background shape for various triggers.
           For each, the acceptance boundary is given by the point at which the distribution has reach
           its maximum and it can no longer be described as a steeply falling background.}
  \label{fig:BkgrShapeAllTrig} 
\end{figure}

It is natural to expect that for a very large integrated luminosity the $H \rightarrow b\bar{b}$ process
would eventually become significant over the statistical uncertainty of a continuous background.
This strategy, however, has not been pursued by experimental collaborations in part due to the limited
rate of triggered events that can be recorded. Fig.~\ref{fig:TrigEff} shows the efficiency of the most akin 
triggers currently included in active trigger menus.
Individually, none of these triggers goes above a 50\% efficiency,
for the SM Higgs. This efficiency increases for larger masses. When combining events collected via all trigger strategies, 
an 80\% of efficiency for the SM Higgs and nearly a 100\% for very large masses is achieved.
Notice that each trigger yields a different background shape and acceptance boundary. In the combination 
of all triggers, the acceptance boundary is still shifted to 80 \GeV\  therefore limiting the accessibility to $Z\to b\bar{b}$ channel.
This is shown in Fig.~\ref{fig:BkgrShapeAllTrig}.

In this letter, a trigger with two b-tagged jets recoiling against one or more jets is proposed, requiring in addition the \pT\ 
of a b-tagged jet pair to be above 200 \GeV. 

\section{Results}
\label{sec:Results}

\begin{figure}[!b]
  \includegraphics[width=0.49\textwidth]{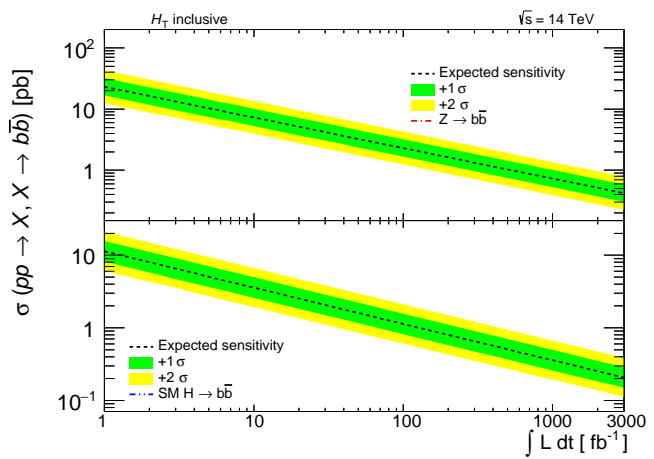}
  \caption{Expected sensitivity to SM $Z$-boson (top panel) and Higgs (bottom panel) inclusive production in the $X$+jets topology.
    This is presented against the integrated luminosity. For comparison, the SM cross-section prediction for these
     processes is also shown.}
  \label{fig:Limits_Resolved} 
\end{figure}

The boson-candidates mass distribution is analysed in a maximum-likelihood fit using HistFitter~\cite{Baak:2014wma}.
Fig.~\ref{fig:Limits_Resolved} shows the expected 95\% CL limit on the cross-section times branching 
ratio of a Higgs or Z  bosons decaying into $b\bar{b}$, $\mathcal{BR}(X\rightarrow b\bar{b})$.
For $Z$ bosons, with only 1 fb$^{-1}$, sensitivity to cross-sections around 20 pb is observed.
For the SM Higgs and 100 fb$^{-1}$, two-sigma sensitivity to a cross-section of around 2 pb is observed.
This is in agreement with the simple $\mathrm{S}/\sqrt{\mathrm{B}}$ shown in Fig.~\ref{fig:candidatemass}a.
Notice that the early Z-boson observation becomes ideal for an {\it in-situ}  uncertainty constraint to be used
in the later $H\rightarrow b\bar{b}$ measurement.

\subsection{Robustness against systematic uncertainties}

The impact of various potential sources of systematic uncertainties is studied by varying the 
signal and background  rates and shapes. 
The aim is to access the impact of potential extra sources of uncertainties 
that were not accounted so for in the MC generation, as for instance, detector effects.
In HistFitter, each systematic variation is associated to a nuisance parameter
that is fitted simultaneously with the signal strength. Nuisance parameters can be constrained by the data,
corresponding in this case to the MC background prediction. Large uncertainties, as those testing the
background MC modelling, are over-constrained minimising their impact. This undesired feature can be 
avoided by separating uncertainties in two classes: those likely to be encountered in a real data-driven 
background scenario and those that are only relevant when considering the potentially inaccurate description
of the background by the MC. The former are assigned nuisance parameters while to test the impact of the latter
the nominal background prediction is substituted by the systematic variation.

\begin{figure}[!t]
  \includegraphics[width=0.49\textwidth]{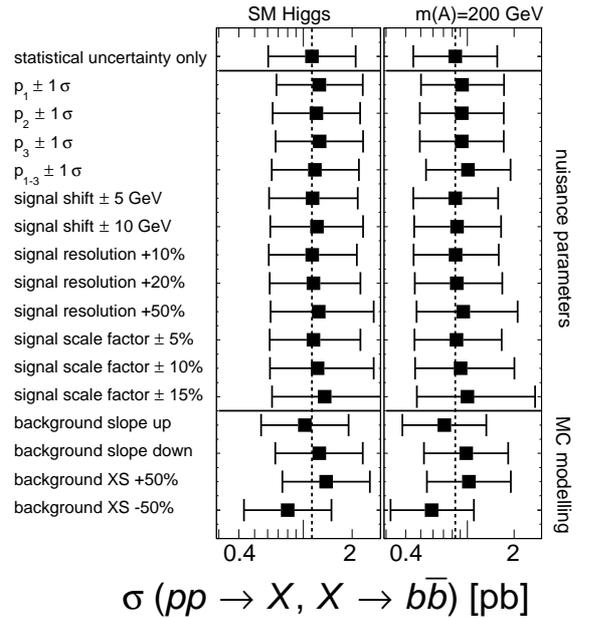}
  \caption{Comparison between the expected sensitivity to the signal cross-section times $\mathcal{BR}(X\rightarrow b\bar{b})$ 
    for the statistical uncertainty only  and the limit after the addition of individual sources of uncertainties.
    This is shown for SM Higgs (left) and pseudo-scalar ${m_A=200}$~\GeV\ (right) hypotheses. The marker shows the
     central value while the error bar corresponds to a $\pm 2 \sigma$ band.}
  \label{fig:Systematics} 
\end{figure}

Fig.~\ref{fig:Systematics} compares the nominal sensitivity to that obtained after including a given uncertainty,
for an integrated luminosity of 100 fb$^{-1}$. The top entry shows the nominal value. The error bars
correspond to $\pm 2 \sigma$. The central panel shows the impact of uncertainties considered as nuisance 
parameters, such as those affecting the signal acceptance and the statistical uncertainty on the parameters of 
Eq.~\ref{eq:FitForm}. From this class, the uncertainty on the signal mass resolution is shown to have the
largest effect, particularly on the $\pm 2 \sigma$ band. The bottom panel shows the impact of uncertainties due
to the MC modelling of the background. Apart from a 50\%\ drop on the MC background cross-section, which as 
expected improves the sensitivity, these uncertainties have a minor impact on the overall result.

\subsection{SM Higgs production cross-section measurement}

\begin{figure}[!t]
  \includegraphics[width=0.49\textwidth]{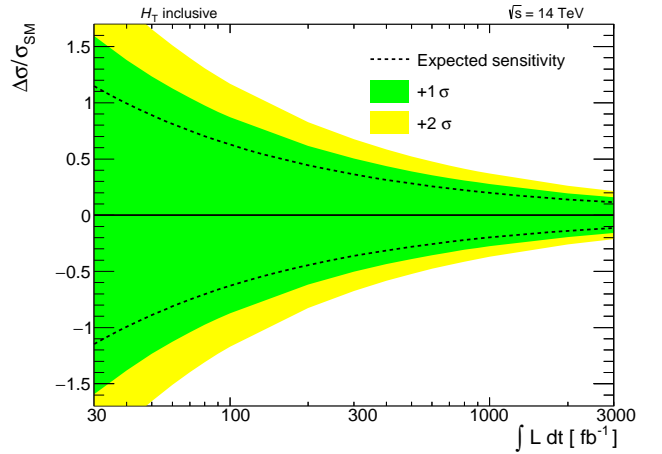}
  \caption{Expected sensitivity to deviations from the SM ${H\rightarrow b \bar{b}}$ cross-section prediction.}
  \label{fig:HiggsCoupling} 
\end{figure}

The sensitivity to deviations from the SM Higgs production cross-section prediction
is tested by assuming the signal hypothesis
\begin{equation}
N_{\mathrm{obs}}=N_{\mathrm{SM\ H\rightarrow b\bar{b}}}+N_{\mathrm{signal}},
\end{equation}
where $N_{\mathrm{signal}}$ is proportional to the fitted parameter $\sigma'$.
Here $N_{\mathrm{obs}}$ is given by the number of expected events from multi-jet, $t\bar{t}$
and SM H$\rightarrow b\bar{b}$. Fig.~\ref{fig:HiggsCoupling} shows the expected sensitivity relative to the 
SM prediction. 
Here, for an integrated luminosity of 100 fb$^{-1}$, a sensitivity to 50\%\ deviations is achieved.
At the high luminosity LHC $\mathcal{L}=3$~ab$^{-1}$  a sensitivity of 10\%\ deviations can be obtained. 

\section{Summary}
\label{sec:summary}

This is the first study to demonstrate a simple and feasible strategy for discovering $X\rightarrow b\bar{b}$
exploring the  continuous and steeply falling background via a suitable fit.
In this letter is shown that the 14~\TeV\ LHC
can observe the $X\rightarrow b\bar{b}$ decay very early for the Z and Higgs bosons. Moreover, the Higgs uncertainties 
for this decay can achieve 10\% for the high luminosity LHC, therefore competitive to the other production channels. 


\begin{table*}[ht]
\centering
\caption{
    Signal and background acceptance measures, presented inclusively and in bins of $H_{\mathrm{T}}$, 
    corresponding to the following ranges in \GeV: 
    400-600; 600-1000; 1000-1600; 1600-2500; 2500 and beyond.  
    For background, the measure shown is the cross-section 
    corresponding to the fraction of events after event selection, including tagging and mistagging efficiencies. 
    This is followed, after the comma, by the lower boundary of the m(b-jet pair) spectrum that can be be described by the function used. 
    The symbol -- is used for bins where either the functional form does not describe the background or the MC statistics are insufficient. 
    For signal, the selection efficiency is shown comma-separated to the fraction of selected events within a 10\% mass window 
    around the generator mass for each sample. 
}
\begin{tabular}[c]{cc cc cc}
\hline
\hline
inclusive & $H_{\mathrm{T}}$ bin 1 & $H_{\mathrm{T}}$ bin 2 &  $H_{\mathrm{T}}$ bin 3 &  $H_{\mathrm{T}}$ bin 4 & $H_{\mathrm{T}}$ bin 5 \\ 
\hline
\multicolumn{6}{c}{Total background cross-section times selection acceptance [pb], acceptance boundary [GeV]} \\ 
\hline
4.9e+02, 60 & 2.8e+02, 60 & 1.6e+02, 90 & 22, 100 & 2.8, -- & 0.21, -- \\ 

\hline
\multicolumn{6}{c}{Signal signal selection efficiency, fraction of selected events within 10\% mass window} \\ 
\hline
\multicolumn{6}{c}{SM $Z$-boson} \\ 
\hline
\ 0.055, 0.72 \ & \ 0.036, 0.72 \ & \ 0.015, 0.7 \ & \ 0.0008, 0.58 \ & \ 3.5e-05, 0.38 \ & \ 9.9e-07, 1.4e-08 \ \\ 

\hline
\multicolumn{6}{c}{SM Higgs-boson} \\ 
\hline
\ 0.1, 0.69 \ & \ 0.052, 0.7 \ & \ 0.04, 0.68 \ & \ 0.0064, 0.72 \ & \ 0.00099, 0.6 \ & \ 0.00036, 0.17 \ \\ 

\hline
\multicolumn{6}{c}{m(A)=70 GeV} \\ 
\hline
\ 0.092, 0.83 \ & \ 0.056, 0.85 \ & \ 0.028, 0.8 \ & \ 0.0018, 0.52 \ & \ 0.00028, 0.073 \ & \ 0.00029, 0.0072 \ \\ 

\hline
\multicolumn{6}{c}{m(A)=200 GeV} \\ 
\hline
\ 0.14, 0.53 \ & \ 0.057, 0.54 \ & \ 0.065, 0.52 \ & \ 0.012, 0.56 \ & \ 0.0029, 0.66 \ & \ 0.00091, 0.54 \ \\ 

\hline
\hline
\end{tabular}
\label{table:table_sample_info}
\end{table*}

\acknowledgments
NG thanks Jon Butterworth, James Ferrando, Andy Pilkington and Michael Spannowsky for useful discussions.
Also, thanks to Dorival Gon\c{c}alves for helping with the generation of the signal samples.
This work is supported by the 7th Framework Programme of the European Commission through
the Initial Training Network HiggsTools PITN-GA-2012-316704.

\bibliographystyle{eplbib}
\bibliography{paper}

\section{Boosting the $b \bar{b}$ system}

To further decrease the event rate, the jet \pT\ threshold could be raised, therefore boosting the $b \bar{b}$ system.
The system boost is parametrised in terms of the scalar sum of all jets passing the event selection $H_{\mathrm{T}}$.
The global observable $H_{\mathrm{T}}$ is chosen instead of the more intuitive $b\bar{b}$ pair \pT,
as it does not require to uniquely identify a candidate,
in line with the inclusive approach presented in this paper.

Table~\ref{table:table_sample_info} summarises the changes on the signal and background acceptance
in terms of the overall boost of system. Here, the total background cross-section drops by one (two)
orders of magnitude for the $3^{\mathrm{rd}}$ ($4^{\mathrm{th}}$) $H_{\mathrm{T}}$ bin, regardless of $R$.
This drop is accompanied by a reduction of the fitting range and of the signal efficiency for
SM $Z$-bosons and hence, from the $2^{\mathrm{nd}}$ $H_{\mathrm{T}}$ bin onward, sensitivity to this process is not expected.
For the lowest mass points, $m_A=70$~\GeV\ and to a lesser extent for SM $Z$-bosons, at very-large $H_{\mathrm{T}}$, 
a considerable loss in signal acceptance and mass resolution is observed.
This loss could be recovered by using $R=0.2$, however, this change is observed to bring little improvement to the other mass points, 
in all but the final $H_{\mathrm{T}}$ bin,
showing that the bulk of the events is in the resolved regime, except for very low mass signals at very-large $H_{\mathrm{T}}$.

\end{document}